\documentclass[%
 reprint,
superscriptaddress,
 amsmath,amssymb,
 aps,
]{revtex4-1}

\usepackage{multirow,graphicx}% Include figure files
\usepackage{booktabs,dcolumn}
\usepackage{bm}% bold math
\usepackage{color}
\usepackage[normalem]{ulem}

\usepackage{diagbox}
\newcolumntype{C}[1]{>{\centering\let\newline\\\arraybackslash\hspace{0pt}}m{#1}}
\AtBeginDocument{
\heavyrulewidth=.08em
\lightrulewidth=.05em
\cmidrulewidth=.03em
\belowrulesep=.65ex
\belowbottomsep=0pt
\aboverulesep=.4ex
\abovetopsep=0pt
\cmidrulesep=\doublerulesep
\cmidrulekern=.5em
\defaultaddspace=.5em
}

\usepackage[colorlinks,citecolor=blue]{hyperref}
\usepackage[colorinlistoftodos]{todonotes}

\begin{document}

\preprint{}

\title{Role of dynamic pairing correlations in fission dynamics}% Force line breaks with \\
%\thanks{}%

\author{R.~Bernard}
\affiliation{%
Centre de mathématiques et de leurs applications, CNRS, ENS Paris-Saclay, Université Paris-Saclay, 94235, Cachan cedex, France.}%
\affiliation{%
CEA, DAM, Île-de-France, 91297 Arpajon, France
}%

\author{S.~A.~Giuliani}
\affiliation{
  NSCL/FRIB Laboratory,
  Michigan State University, 
  East Lansing, Michigan 48824, USA 
}%

\author{L.~M.~Robledo}
\email{lm.robledo.physics@gmail.com}
\affiliation{%
  Center for Computational Simulation,
  Universidad Polit\'ecnica de Madrid,
  Campus de Montegancedo, 
  Boadilla del Monte, 28660-Madrid, Spain
}%
\affiliation{%
  Departamento de F{\'i}sica Te{\'o}rica, 
  Universidad Aut{\'o}noma de Madrid, 
  28049 Madrid, Spain}

\date{\today}% It is always \today, today,
             %  but any date may be explicitly specified

\begin{abstract} 
	We study the role of dynamic pairing correlations in fission dynamics by
	considering intrinsic Hartree-Fock-Bogoliubov wave functions that are
	obtained by minimizing the particle number projected energy.  For the
	restricted variational space, the set of self-consistent wave functions
	with different values of proton and neutron number particle fluctuations
	are considered. The particle number projected energy is used to define
	potential energy surface for fission whereas collective inertias
	are  computed within the traditional formulas for the
	intrinsic states. The results show that the effect of the restricted
	variation after particle number projection in the potential energy
	surface is small while collective inertias
	substantially decrease.  On the other hand,
	we show that this quenching is strongly mitigated when Coulomb
	antipairing is considered. In the light of these beyond mean-field
	calculations, the validity of traditional fission calculations is
	discussed. 
\end{abstract}

 %PACS, the Physics and Astronomy
                             % Classification Scheme.
%\keywords{Suggested keywords}%Use showkeys class option if keyword
                              %display desired
\maketitle

%\tableofcontents

\section{\label{sec:level1}Introduction}

Undoubtedly, pairing correlations represent a key ingredient in the description
of the dynamics of the fission phenomenon experienced by heavy atomic
nuclei~\cite{Moretto1974,Schunck2016}. For instance, the amount of pairing
correlations has a strong impact on quantities such as spontaneous fission
lifetimes
\cite{Urin1966,Staszczak1985,Lazarev1987a,Staszczak1989,Lojewski1999,Pomorski2007,Mirea2010,Giuliani2013,Rodriguez-Guzman2014,Giuliani2014,Zhao2016},
the shape of the barriers separating the ground state from
scission~\cite{Stepien1968,Rutz1999,Samyn2005,Schunck2014,Abusara2010,Karatzikos2010,Sadhukhan2014}
and fission fragments
distributions~\cite{Goutte2005,Sadhukhan2015,Tao2017,Matheson2018}.
At the mean-field level, pairing is traditionally described using the
Hartree-Fock-Bogoliubv (HFB) theory, which is a reasonable approximate scheme
when pairing correlations are strong~\cite{ring1980}. In nuclear physics,
however, the pairing strength is not strong enough and, as a consequence, many
mean field configurations show little or no pairing
correlations~\cite{Dean2003,Brink2005}. In this case the mean-field description
of the nucleus breaks down, and the inclusion of dynamic pairing correlations
stemming from beyond mean-field effects becomes necessary. The evolution of the
nucleus through the different shapes involved in fission affects the level
density around the Fermi energy, with a large impact on pairing correlations.
This effect is reflected by the intricate behavior shown along the fission path,
including many regions of very weak static pairing which points out the possible
crucial role of dynamic pairing correlations in the studies of fission.

In order to account for such effects, beyond mean field calculations involving
the restoration of the particle number of the nuclear wave function are
required.  Unfortunately the computational cost of beyond mean field
calculations limited so far their application to fission studies, keeping the
impact of dynamic pairing correlations unexplored.  Moreover, to properly
address the role of dynamic pairing correlation one should account for all those
effects that may mitigate the effective pairing strength and that are usually
neglected for the sake of computational time, like for instance Coulomb
antiparing~\cite{Anguiano2001b}, which is the name given to the destructive
effect of the repulsive Coulomb interaction in proton's pairing correlations. If
proton and neutron pairing strengths are independently adjusted to experimental
data in the region of interest~\cite{Bertsch2009}, Coulomb anti-pairing is taken
into account in an effective way by the fitted pairing strengths.  Conversely,
in forces like Gogny~\cite{decharge1980} the neutron pairing strength is fitted
to experimental data (for instance in the tin isotopic chain) and the proton
pairing strength comes from isospin invariance.  In those cases, Coulomb
anti-pairing must be explicitly taken into account to avoid the self-energy
problem and the breaking of the Pauli principle in particle number projected
calculations. The Coulomb anti-pairing effect can reduce proton's
pairing gap by a $20-30$\%~\cite{Lesinski2009,Nakada2011}, with a strong impact
on observables such as moments of inertia~\cite{Anguiano2001b,Afanasjev2014}, but
their effect is usually neglected  due to the enormous computational cost associated to the evaluation
of Coulomb's pairing field~\cite{Anguiano2001b}. 

In the light of this discussion, it is possible to conclude that the
inclusion of dynamic pairing will have a twofold effect:
On the one hand, collective inertias driving fission dynamics, with their
inverse dependence on the square of the pairing
gap~\cite{brack1972,Moretto1974,Bertsch1991,Giuliani2014}, are expected to
increase when the Coulomb anti-pairing effect is considered, increasing the
collective action and leading to longer fission lifetimes $t_\mathrm{SF}$. On
the other hand, dynamic pairing correlations are expected to increase the
pairing gap reducing thereby the collective inertias.  The outcome of these
competing effects is uncertain and it is the purpose of this paper to clarify
this situation and establish a step forward in the study of beyond mean-field
effects in fission calculations. 

In previous studies, angular momentum projection~\cite{Bender2004} has
been used to compute fission barrier heights. However, the results are almost
indistinguishable from the ones obtained with the rotational
correction~\cite{Schunck2016,Rodriguez-Guzman2000}. Parity projection has also
been considered in the reflection asymmetric section of the fission
path~\cite{Hao2012,Samyn2005} with little or no impact at all. Finally, the
impact of particle number projection on fission barrier heights has been
considered in~\cite{Samyn2005}. A change of at most $\pm$ 0.5 MeV is obtained in
all the cases. 

In this paper we are considering the
contribution to dynamic pairing correlation coming from a restricted variation after
projection for particle number projection. The evolution of the pairing
properties of the nucleus as it evolves towards fission, will be studied as a
function of the axial quadrupole moment $q=\langle Q_{20}\rangle$. We will
analyze the impact of dynamic pairing correlations in the potential energy
surface, computed with the particle number projected wave function
$|\Psi^N(q)\rangle=\hat P^N|\varphi(q)\rangle$, and in the collective inertia
computed with the intrinsic state $|\varphi (q)\rangle$ associated to the
former.  

% ----------------------------------------------------------------------------------
\section{Methodology}

\begin{figure}
    \centering
    \includegraphics[width=11cm]{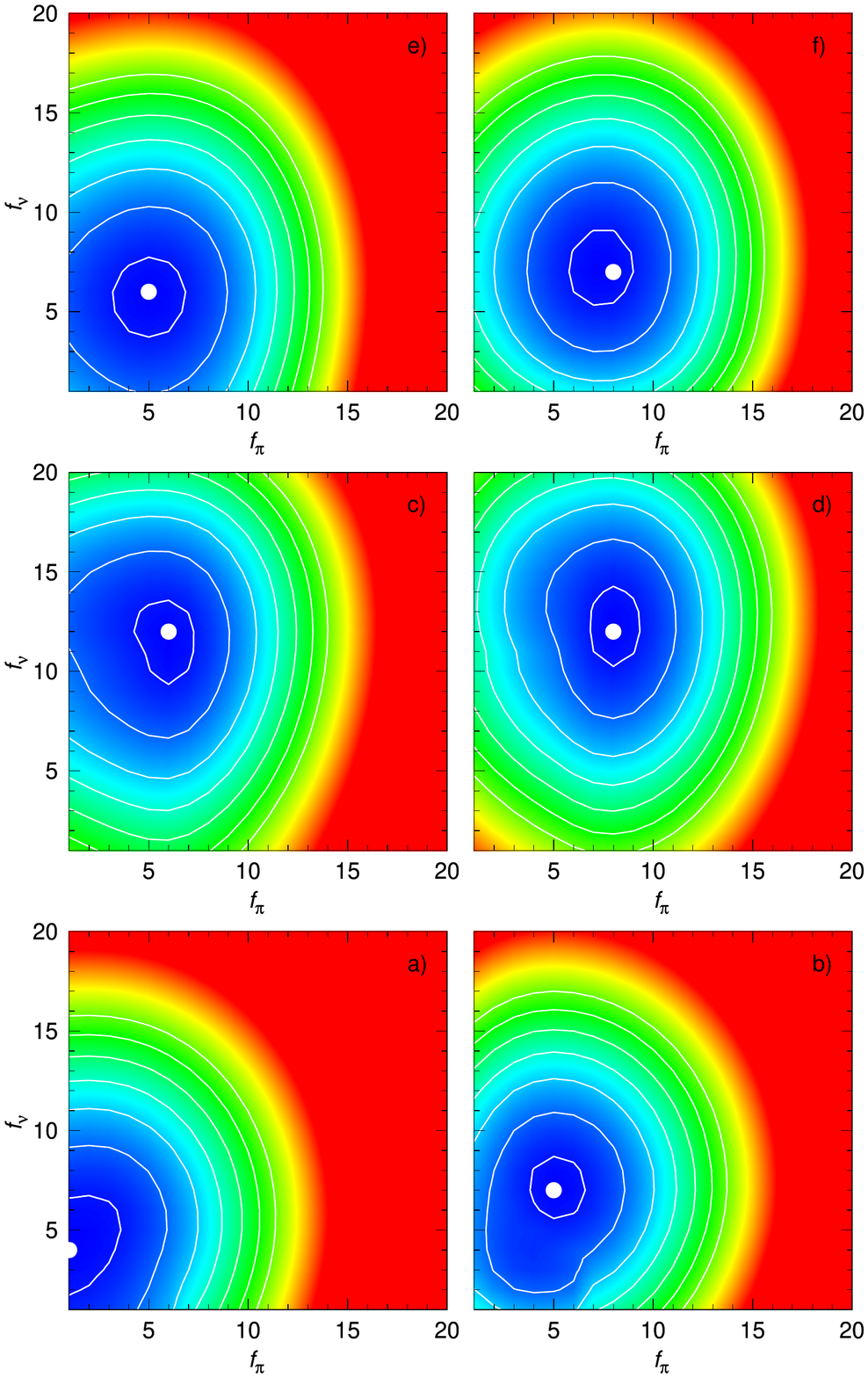}
    \caption{(Color online) In the left (right) panels contour plots of the HFB
    (PNP) energy as a function of $\langle \Delta N^2\rangle $ and $\langle
    \Delta Z^2\rangle $ are given for three different quadrupole moments.
    Namely, $Q_2=14$b (ground state) (panels a and b), $Q_2=28$b (first fission
    barrier) (panels c and d), and $Q_2=42$b (fission isomer) (panels e and f).
    The results are obtained with the Gogny D1M force for the nucleus
    $^{240}$Pu. The minima are marked by a dot and the color range spans 5 MeV.}
    \label{fig:240PuDN2}
\end{figure}

Dynamic pairing correlations require a beyond mean field framework involving the
restoration of the particle  quantum number of the nuclear wave function.
In order to gain more correlations, the intrinsic mean field wave function has
to be determined by minimizing the projected energy in the so called variation
after projection (VAP) method.  In this paper we use the restricted variation
after projection (RVAP)~\cite{Rodriguez2005} particle number projection (PNP)
method~\cite{Anguiano2001}.  The RVAP-PNP has been shown to be superior to other
alternatives like the Lipkin-Nogami method~\cite{Rodriguez2005} commonly used in
the literature.  In the RVAP-PNP method the variational subspace is formed by
projecting onto good particle number (protons and neutrons separately) intrinsic
wave functions obtained from a  HFB calculation constraining on the particle
number fluctuation for protons and neutrons  $|\Phi (\langle \Delta
N^2\rangle_\pi,\langle \Delta N^2\rangle_\nu)\rangle$~\footnote{The constraint
on $\protect \langle \Delta N^2 \protect\rangle$ involves the two-body operator
$\protect\Delta \protect\hat{N}^2$. The implementation of this constraint is
straightforward when using the gradient method to solve the HFB
equation~\cite{Robledo2011}.}. Henceforth, we will denote by $f_\pi$ and $f_n$
the value of the particle number fluctuation for protons and neutrons,
respectively.  The RVAP intrinsic state $|\Phi (f_\pi,f_\nu) \rangle$
corresponds to the
minimum of the projected energy
\begin{equation}
    E^{Z,\,N} (f_\pi,f_\nu) = \frac
{
    \langle \Phi (f_\pi,f_\nu) | \hat{H} P^Z P^N |
    \Phi (f_\pi,f_\nu) \rangle
}
{    \langle \Phi (f_\pi,f_\nu) |
    \Phi (f_\pi,f_\nu) \rangle
} \,,
\end{equation}
as a function of the $f_\pi$ and $f_\nu$ variables. The minimum of the two
dimensional function $E^{Z,\,N} (f_\pi,f_\nu)$ is determined by a simple
gradient method in two dimensions.  The potential energy surface for fission is
obtained by introducing an additional constraint on the quadrupole moment
$Q_{20}$ of the axially symmetric intrinsic state and is given by the projected
energy of the RVAP for each $Q_{20}$ value. We could also introduce easily
additional constrains like the octupole moment or the necking operator to form
multidimensional potential energy surfaces (PES) so popular in fission studies,
but this is not the purpose of the present work. An example of both the HFB and
PNP potential energy surfaces obtained as a function of $\langle \Delta
N^2\rangle $ and $\langle \Delta Z^2\rangle $ is given in
Fig.~\ref{fig:240PuDN2} where those energies, computed with the Gogny D1M
parametrization~\cite{goriely2009}, are plotted for the nucleus
$^{240}$Pu and three different values of the quadrupole moment (see caption for
details). The chosen quadrupole moments correspond to the ground state, first
fission barrier and fission isomer.  Both the HFB and PNP energies show a 
parabolic behaviour as a function of $(f_\pi,f_\nu) $ that is
slightly distorted in both cases. In the figure, it is clearly observed how the
minimum of the PNP energy is shifted to higher $\langle \Delta N^2\rangle $ and
$\langle \Delta Z^2\rangle $ values as compared to the  HFB ones. This is in
agreement with the expectation that the RVAP method provides intrinsic
states with more pairing correlations than those intrinsic states obtained 
by the HFB method. This has important consequences for fission dynamics as
the collective inertia strongly depend upon the amount of pairing correlations.

The other quantity required to study the dynamics of spontaneous fission is the
collective inertia associated to the collective variables used to drive the
nucleus from its ground state to fission. The collective inertia plays a crucial
role in several fission observables, such as the spontaneous fission lifetimes
$t_\mathrm{SF}$ obtained within the Wenzel-Kramers-Brillouin (WKB) formula and
the fission fragments distributions obtained in both time dependent frameworks~\cite{Goutte2005,Regnier2016} and stochastic Langevin
approaches~\cite{Sadhukhan2015,Matheson2018}.  For instance, the $t_\mathrm{SF}$ has an
exponential dependence on the collective inertia than can amount to changes of
several orders of magnitude in this quantity~\cite{Giuliani2013,Giuliani2014,Rodriguez-Guzman2014}. As mentioned before, the magnitude of
the collective inertia depends on the amount of pairing correlations in a way
that can be quantified as an inverse dependence on the square of the pairing gap. 
This dependence on the
amount of pairing correlations implies that the larger the pairing correlations
are the smaller the collective inertia (and therefore $t_\mathrm{SF}$) is.
Therefore, we expect a strong dependence of the collective inertia on the
combined action of both the Coulomb antipairing effect and the PNP\@.  

There are two types of collective inertias: the one coming from adiabatic time
dependent Hartree-Fock-Bogoliubov (ATDHFB) theory and the one coming for the
Gaussian overlap approximation (GOA) to the Generator Coordinate Method 
(GCM)~\cite{Schunck2016}. Unfortunately, so far none of these schemes has been
generalized to the case of non-HFB states like the PNP ones considered in this
paper. In these respect, the GCM-GOA framework is more promising since its
formalism is not intimately rooted to the HFB method. However, the perturbative
cranking approximation (where the linear response matrix is approximated by its
diagonal both in the expressions of the inertia and in the definition of the
collective momentum~\cite{Giuliani2018}) required to alleviate the computational
cost of the collective inertias is not easy to implement in the PNP case.
Therefore we take a pragmatic approach and use for the PNP case the perturbative
cranking inertias computed with the intrinsic state $|\Phi\rangle$ obtained in
the RVAP\@. Work to obtain a sound and easy way to compute the inertia for PNP
wave functions is underway and will be reported elsewhere.

To avoid the appearance of divergences in the calculation of the PNP energy with
the Gogny force, we computed the exchange, direct and pairing channels for each
of the terms of the interaction~\cite{Anguiano2001}.  The required Hamiltonian
and norm overlap between the HFB state $|\Phi\rangle$ and its rotated in gauge
space $e^{i\Phi_p \hat Z}e^{i\Phi_n\hat N}|\Phi\rangle$ are computed using the
methodology of the generalized Wick theorem as developed
in~\cite{Robledo2009,Bertsch2012}. For the density dependent part of the
interaction we use the so-called "PNP projected density prescription" that is
commonly used for particle number projection~\cite{Anguiano2001,Robledo2007} (be
aware, however, of the fundamental difficulties encountered when using  the
projected density prescriptions in the context of spatial symmetries
restoration~\cite{Robledo2010}).

% ----------------------------------------------------------------------------
%
%     S e c t i o n    R e s u l t s
%
% ----------------------------------------------------------------------------
\section{Results}

\begin{figure}
    \centering
    \includegraphics[width=9cm]{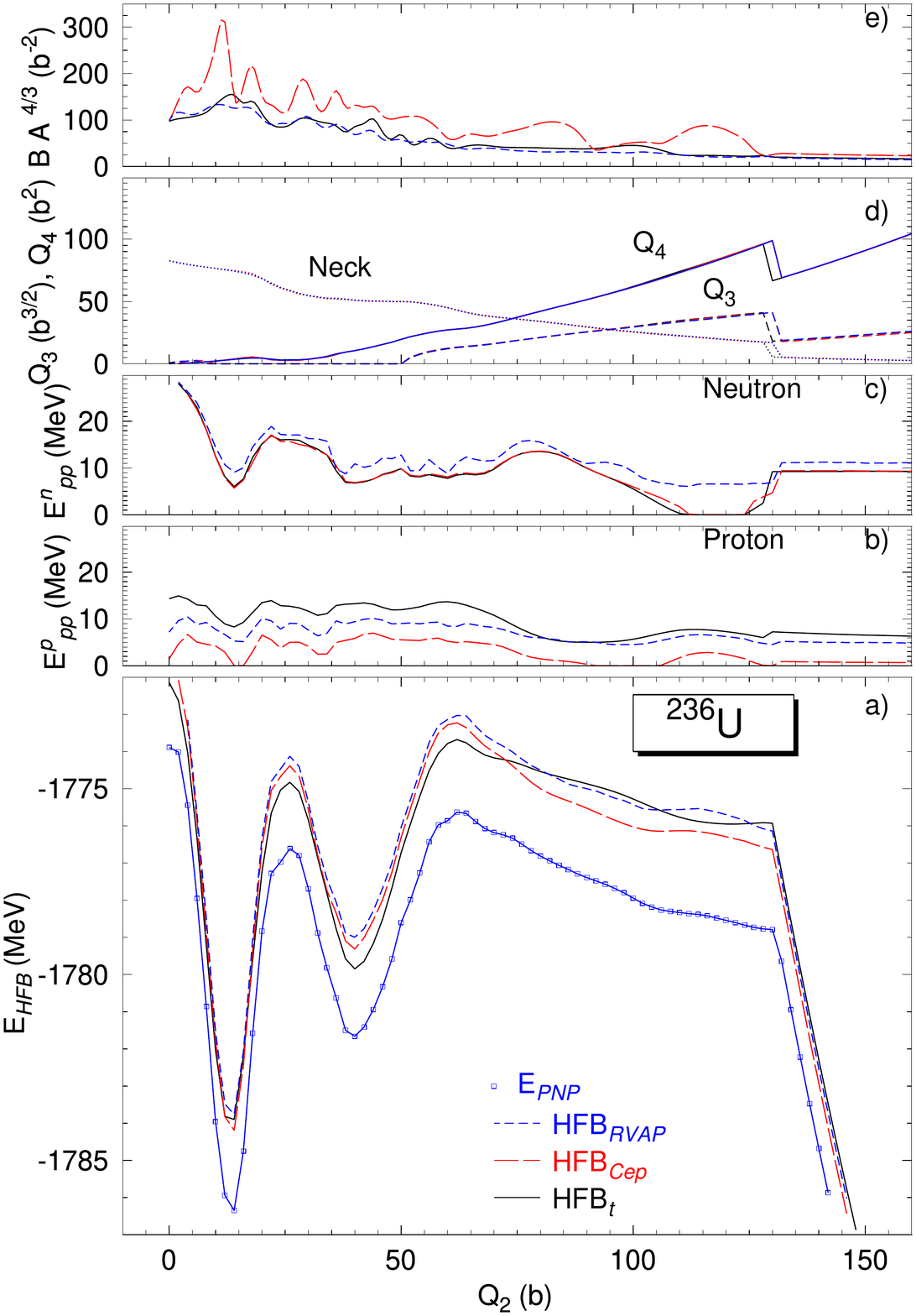}
    \caption{(Color online) In panel a) the potential energies obtained in the
    different approaches discussed in the text are plotted as as function of the
    quadrupole moment of the intrinsic state. The color code and the labels are
    described in the main text.  In panels b) and c) the particle-particle
    correlation energy $\frac{1}{2}\mathrm{Tr} \Delta_\tau \kappa_\tau$ for
    protons and neutrons, respectively, is given. The octupole, hexadecapole and
    neck parameters are given in panel d). Finally, in panel e) the ATDHFB
    quadrupole collective inertia computed in the perturbative approximation is
    given.
    \label{fig:236U}}
\end{figure}

We have considered three nuclei as prototipical examples of application of the
issues discussed in the previous section.  The first nucleus studied is the
light actinide $^{236}$U, characterized by a double humped potential energy
surface (PES) with high and wide barriers.  Reflection symmetry is broken right
after the fission isomer and therefore asymmetric fragment mass distribution is
expected for this nucleus. In Fig.~\ref{fig:236U} we show the most relevant
quantities for a theoretical understanding of fission. In panel
\hyperref[fig:236U]{a)}  potential energy surfaces (to be discussed below) are
shown as a function of the quadrupole moment. The corresponding
particle-particle correlation energies $\frac{1}{2}\mathrm{Tr} ( \Delta_\tau
\kappa_\tau)$ (with $\tau=p,n$) are shown in panels \hyperref[fig:236U]{b)} and
\hyperref[fig:236U]{c)}. In panel \hyperref[fig:236U]{d)} the self-consistent
octupole and hexadecapole moments are also shown along with the neck parameter
given by the mean value of the neck operator $Q_N=\exp [-(z-z_0)^2/a_0^2]$ with
$z_0=0$ and $a_0=1.0$ fm.  Finally, in panel \hyperref[fig:236U]{e)} the
collective inertia computed in the traditional perturbative ATDHFB scheme is
displayed.  

Panel \hyperref[fig:236U]{a)} shows the potential energy surfaces for four different
calculations. The black solid line (HFB$_t$) corresponds to the traditional HFB
calculation where Coulomb exchange is evaluated in the Slater approximation and
Coulomb and spin-orbit anti-pairing are neglected. The dashed red line
(HFB$_\mathrm{Cep}$) corresponds to a HFB calculation where both Coulomb
exchange and anti-paring are fully considered. Comparing the predicted isomer
energies ($E_{II}$) and inner ($B_{I}$) and outer ($B_{II}$) fission barrier
heights (see Table~\ref{table1}) we notice that HFB$_\mathrm{Cep}$ predicts
values that are $0.75-0.83$~MeV larger. This increase is an expected behavior
when pairing correlations get reduced~\cite{Samyn2005,Rodriguez-Guzman2014}. Also, more
pronounced structures are observed in HFB$_\mathrm{Cep}$, particularly at large
quadrupole deformations, which can be traced back to the reduced pairing
correlations~\cite{Bertsch2018} associated to the presence of Coulomb
anti-pairing. These changes in the potential energy surface are partially washed
out in the HFB calculation obtained with intrinsic RVAP states
(HFB$_\mathrm{RVAP}$, blue dashed line). The HFB$_\mathrm{RVAP}$ barriers
heights and isomer excitation energy are $0.52-0.55$ larger than the HFB$_t$,
and the potential energy surfaces at large deformations are also similar. This
result suggests that pairing correlations induced by the RVAP partially cancel
out the effect of the Coulomb anti-pairing quenching (see below). Finally, the
blue full curve with symbols corresponds to the RVAP projected energy
($E_\mathrm{PNP}$). This energy is around two MeV deeper than the intrinsic
energies, being the fission parameters $0.50-0.64$~MeV larger than the HFB$_t$
results.

In order to better understand the impact of dynamic correlations on fission, it
is worth to analyze the changes in the other quantities depicted in
Figure~\ref{fig:236U}. Proton particle-particle correlation energies are shown
in panel \hyperref[fig:236U]{b)} for the HFB$_t$, HFB$_\mathrm{Cep}$ and
HFB$_\mathrm{RVAP}$ intrinsic states (this quantity is meaningless in the PNP
case). Coulomb antipairing quenches the particle-particle proton correlation
energy, but the quenching is softened by the effect of the PNP-RVAP, being the
latter results closer to the HFB$_t$ ones. In the neutron case, shown in panel
\hyperref[fig:236U]{c)}, no significant differences are observed between the
HFB$_t$ and HFB$_\mathrm{Cep}$ cases as expected. The effect of PNP-RVAP is to
increase neutron pairing correlations bringing the particle-particle correlation
energy of the intrinsic state above the other two curves. The quadrupole,
octupole and necking shape parameters are shown in panel
\hyperref[fig:236U]{d)}. For each of the  three parameters, the results obtained
with the three different types of intrinsic states lie each on top of the other.
The impact of different types of pairing correlation regimes on the deformation
parameters is negligible. Finally, in panel \hyperref[fig:236U]{e)} the ATDHFB
perturbative collective inertia for the three intrinsic states are shown. As
compared to the HFB$_t$ reference calculation, the HFB$_\mathrm{Cep}$ inertia is
larger as a consequence of the quenched pairing.  Overall, the
HFB$_\mathrm{Cep}$ inertia is around two times larger than the HFB$_t$ one. It
also shows more pronounced structures in the form of high peaks.  On the other
hand, the increase of pairing correlations associated to PNP-RVAP brings the
HFB$_\mathrm{RVAP}$ intrinsic inertia back to the range of the HFB$_t$ curve. It
is worth mentioning that the HFB$_\mathrm{RVAP}$ inertia looks a bit smoother
than the HFB$_{t}$ one.  From this comparison we conclude that the HFB$_t$
inertia (i.e.\ without Coulomb exchange, and what is more important, without
Coulomb anti-pairing) represents a good approximation, in the case of the Gogny
force, to the inertia obtained from the PNP-RVAP intrinsic states. It is worth
mentioning that this cancellation is typical of the Gogny forces and is not
expected in calculations where the strength of the pairing interaction is fitted
separately for protons and neutrons to experimental data~\cite{Baran2015}. In
this case, the effect of Coulomb anti-pairing is taken into account by the
fitted pairing strength and therefore a reduction of a factor of two in the
inertias has to be expected.

Finally, we have computed the spontaneous fission half-live $t_\mathrm{SF}$
using the traditional WKB formula (see
Refs~\cite{Schunck2016,Rodriguez-Guzman2014} for details and applications) with
a $E_0$ parameter of 1 MeV. The results for the HFB$_t$ and HFB$_\mathrm{Cep}$
cases are computed with the corresponding PES and collective inertias, whereas
the PNP-RVAP is computed with the PNP PES but using the collective inertia of
the HFB$_\mathrm{RVAP}$ intrinsic state. The results are summarized in
Table~\ref{table1} along with the values of $E_{II}$, $B_{I}$ and $B_{II}$
discussed above. The effect of Coulomb antipairing in the inertia increases
$t_\mathrm{SF}$ by 20 (14) orders of magnitude in the ATDHFB (GCM) cases, but
this huge impact is cancelled out by the dynamic pairing effect associated to
RVAP-PNP\@. The final RVAP-PNP $t_\mathrm{SF}$  values are very close to the
HFB$_{t}$ ones. It is important to emphasize that the RVAP-PNP  $t_\mathrm{SF}$
values are lower than the HFB$_{t}$ ones in spite of the larger fission barrier
heights. This is due to the smaller values of the inertias in the projected
case.

\begin{table}

\begin{tabular}{clccccc}

\toprule

&		    &  \multicolumn{2}{c}{t$_\mathrm{SF}$}        & \\
		                \cmidrule(r){3-4}                 
&		    &  ATDHFB              &  GCM                 &  B$_{\textrm{I}}$ & E$_{\textrm{II}}$ & B$_{\textrm{II}}$ \\
&		    &  (s)                 &  (s)                 &  (MeV)         &  (MeV)         &  (MeV)         \\

\midrule

\parbox[t]{3mm}{\multirow{4}{*}{\rotatebox[origin=c]{90}{$^{236}$U}}} 

& \multicolumn{1}{|l}{HFB$_{t}$}           &  $3.0\times10^{43}$  &  $2.4\times10^{32}$  &  9.07  &  4.05  &  10.22  \\
& \multicolumn{1}{|l}{HFB$_\mathrm{Cep}$}  &  $3.1\times10^{63}$  &  $1.2\times10^{46}$  &  9.82  &  4.88  &  10.97  \\
& \multicolumn{1}{|l}{HFB$_\mathrm{RVAP}$} &  $8.3\times10^{41}$  &  $1.1\times10^{32}$  &  9.64  &  4.77  &  10.74  \\
& \multicolumn{1}{|l}{PNP}                 &  $1.0\times10^{42}$  &  $1.4\times10^{32}$  &  9.74  &  4.69  &  10.72  \\

\midrule

\parbox[t]{3mm}{\multirow{4}{*}{\rotatebox[origin=c]{90}{$^{240}$Pu}}} 

& \multicolumn{1}{|l}{HFB$_{t}$}           &  $7.4\times10^{38}$  &  $7.5\times10^{29}$  &  10.23  &  4.39  &  10.20  \\
& \multicolumn{1}{|l}{HFB$_\mathrm{Cep}$}  &  $2.0\times10^{54}$  &  $2.4\times10^{39}$  &  10.91  &  4.94  &  10.75  \\
& \multicolumn{1}{|l}{HFB$_\mathrm{RVAP}$} &  $3.0\times10^{37}$  &  $9.5\times10^{28}$  &  10.74  &  4.74  &  10.57  \\
& \multicolumn{1}{|l}{PNP}                 &  $2.8\times10^{37}$  &  $1.2\times10^{29}$  &  10.83  &  4.79  &  10.63  \\

\midrule

\parbox[t]{3mm}{\multirow{4}{*}{\rotatebox[origin=c]{90}{$^{252}$Cf}}} 

& \multicolumn{1}{|l}{HFB$_{t}$}           &  $2.3\times10^{22}$  &  $1.7\times10^{18}$  & 11.18   &  3.71   &   7.77  \\
& \multicolumn{1}{|l}{HFB$_\mathrm{Cep}$}  &  $7.6\times10^{24}$  &  $2.9\times10^{18}$  & 11.60   &  3.45   &   6.86  \\
& \multicolumn{1}{|l}{HFB$_\mathrm{RVAP}$} &  $7.8\times10^{19}$  &  $2.5\times10^{15}$  & 11.19   &  3.40   &   7.09  \\
& \multicolumn{1}{|l}{PNP}                 &  $1.9\times10^{21}$  &  $6.2\times10^{16}$  & 11.22   &  3.71   &   7.49  \\

\bottomrule

\end{tabular}

\caption{On the left hand side, spontaneous fission half-lives (in seconds) computed with two
different sets of collective inertias (ATDHFB and GCM) and for the four
different sets of calculations for the nuclei considered. On the right hand
side, the values of 
the fission barrier heights $B_I$, $B_{II}$ and fission isomer excitation energy $E_{II}$ (in MeV)
are algo given.}
\label{table1}	

\end{table}

The results obtained for the nucleus $^{240}$Pu look qualitatively the same as
those obtained for $^{236}$U, being the small differences observed mostly due to
shell effects associated with the different proton and neutron numbers.  The
values of $E_{II}$, $B_{I}$ and $B_{II}$ are given in Table~\ref{table1}. The
most notorious difference is in the larger values of $B_{I}$ which are around 1
MeV higher than in the $^{236}$~U case.  The impact of Coulomb antipairing in
$t_\mathrm{SF}$ is 16 (10) orders of magnitude the ATDHFB (GCM) inertias and, as
in the uranium case, the inclusion of dynamical pairing correlations reduce
substantially $t_\mathrm{SF}$ and brings it closer to the traditional HFB$_t$
value. As in the previous case, we conclude that the dynamic pairing compensates
the Coulomb antipairing effect and the $t_\mathrm{SF}$ values obtained in the
traditional HFB approach are very similar to the ones obtained in the RVAP-PNP
context.

\begin{figure}
    \centering
    \includegraphics[width=9cm]{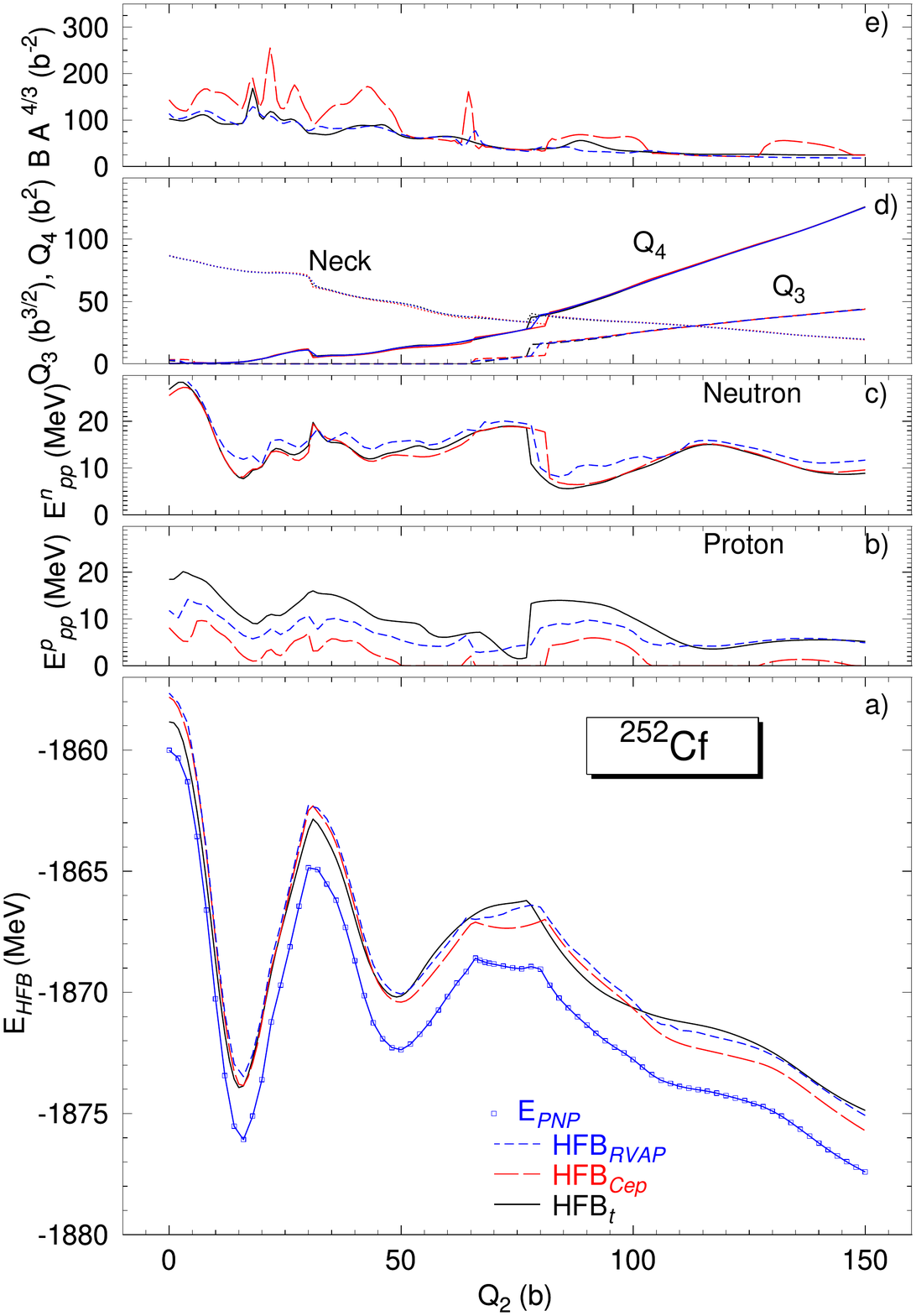}
    \caption{Same as Fig.~\ref{fig:236U} but for the nucleus $^{252}$Cf.
}
    \label{fig:252Cf}
\end{figure}

We have also carried out calculations for the heavier $^{252}$Cf isotope. The
potential energy surfaces, particle particle energy correlations, deformation
parameters and ATDHFB collective inertias are shown in Fig.~\ref{fig:252Cf}. In
all the cases, the PES show a rather high inner barrier (see Table~\ref{table1}
for the values of the different parameters). The reflection symmetric fission
isomer lies at around 3.7 MeV excitation energy, whereas the slightly reflection
asymmetric outer barrier is around 7 MeV high. In this particular nucleus the
impact of the different theoretical schemes used in the calculation of the outer
barrier is stronger with changes in its height of more than 1 MeV.  It turns out
that in the region of the outer barrier the HFB$_{t}$ PES is very flat with
several coexisting minima but one of them is clearly favored  when Coulomb
antipairing is considered.  The particle-particle correlation energy for protons
looks similar to the one of $^{236}$U for the HFB$_\mathrm{Cep}$ and
HFB$_\mathrm{RVAP}$ cases but differs significantly in the HFB$_{t}$ case around
the outer barrier region.  The reason for this behavior is the same that explain
the discrepancies in the PESs in that region.  The $E_{pp}$ for neutrons follows
the same pattern as in the uranium case and only small differences are noticed
in the outer barrier region.  The same observation is valid for the deformation
parameters of panel \hyperref[fig:252Cf]{d)}. The behavior of the ATDHFB inertia
in panel \hyperref[fig:236U]{e)} is qualitatively similar to the one of
$^{236}$U.

Concerning $t_\mathrm{SF}$, we observe longer values when Coulomb antipairing is
considered but the difference amounts to 2 (0) orders of magnitude in the ATDHFB
(GCM) case. This is in strong contrast with the $^{236}$U and $^{240}$Pu cases.
A possible explanation is the reduction of the outer barrier height of more than 1
MeV seen in this particular case. Considering dynamical pairing
lowers $t_\mathrm{SF}$ by 5 (3) orders of magnitude in the ATDHFB (GCM) cases as
compared to the HFB$_\mathrm{Cep}$ result. The net effect of this opposite
trends is to yield final values for the RVAP-PNP calculation which are, again,
pretty close to the HFB$_{t}$ ones.

\section{Conclusions} 

In this paper we studied the
impact of dynamical pairing correlations in the theoretical estimation of
fission properties. We found that particle number projection in the restricted
variation after projection framework (using $\langle\Delta N^2\rangle$ for
protons and neutrons as variational parameters) has a profound impact on some of
the quantities related to fission such as spontaneous fission half-lives.
The
parameters defining the potential energy surface, like the fission barrier
heights and fission isomer location are little affected by particle number
projection in the three examples analyzed. On the other hand, the increase in
pairing correlations due to particle number restoration leads to a quenching of
the collective inertia by a factor of the order of two. The consequences for the
spontaneous fission half-life depend on the nucleus but it is quantified to be
large and can reach a reduction of 20 orders of magnitude. This reduction is
compensated by the Coulomb anti-pairing effect, which is
often neglected (or
described with the Slater approximation) in mean field calculations but
is
required in particle number projection to avoid the self-energy and self-pairing
problems. The reduction of pairing correlations associated with  Coulomb
anti-pairing increases the collective inertias by a factor of around two in the
examples considered and can increase the calculated $t_\mathrm{SF}$  up to 20
orders of magnitude. On the other hand, the consequences of an exact treatment
of the Coulomb potential in the potential energy surface are relatively small
and have a relatively less important impact on $t_\mathrm{SF}$. The two opposite
effects, Coulomb anti-paring and dynamical pairing correlations tend to suppress
each other and the final outcome turns out to be similar to the results obtained
omitting both of them. This result is relevant for calculations with nuclear
forces (Gogny among them), where the nuclear pairing interaction is isospin
invariant and Coulomb anti-pairing has to be considered. The effect of dynamical
pairing correlations alone is relevant for other interactions  where the pairing
strength for protons and neutrons is fitted separately to experimental data.

For future work, the evaluation of the collective inertias with particle
number projected wave functions is the next step to consider. Also, the
consequences of particle number projection on induced fission half-lives
and properties of the fission fragments could be an interesting subject
of research.

\begin{acknowledgments}
  SAG  acknowledge support from the U.S. Department of Energy under Award Number DOE-DE-NA0002847 
  (NNSA, the Stewardship Science Academic Alliances program). The work of
  LMR has been supported in part by Spanish grant Nos FIS2015-63770 MINECO 
  and FPA2015-65929 MINECO.
\end{acknowledgments}

%\bibliography{FissionPNP}
%merlin.mbs apsrev4-1.bst 2010-07-25 4.21a (PWD, AO, DPC) hacked
%Control: key (0)
%Control: author (8) initials jnrlst
%Control: editor formatted (1) identically to author
%Control: production of article title (-1) disabled
%Control: page (0) single
%Control: year (1) truncated
%Control: production of eprint (0) enabled
%

\end{document}